\documentclass[runningheads,a4paper]{llncs}
\usepackage{amssymb} 
\usepackage{graphicx}
\usepackage{url}
\usepackage{float}

\newcommand{\jolkw}[1]{\textbf{\bfseries #1}}
\newcommand{\jol}[1]{\texttt{#1}}

\begin{document}

\title{Microservices Science and Engineering}

\author{Manuel Mazzara \inst{1} \and Kevin Khanda\inst{1} \and Ruslan Mustafin \inst{1} \\ Victor Rivera\inst{1} \and Larisa Safina\inst{1} \and Alberto Sillitti  \inst{1}}
\authorrunning{Manuel Mazzara et al.}

\institute{
Innopolis University, Russian Federation\\
\email{\{m.mazzara,k.khanda, r.mustafin, v.rivera, l.safina, a.sillitti\}@innopolis.ru}
}

\toctitle{Lecture Notes in Computer Science}
\tocauthor{Authors' Instructions}
\maketitle

\begin{abstract}
In this paper we offer an overview on the topic of Microservices Science and Engineering (MSE) and we provide a collection of bibliographic references and links relevant to understand an emerging field. We try to clarify some misunderstandings related to microservices and Service-Oriented Architectures, and we also describe projects and applications our team have been working on in the recent past, both regarding programming languages construction and intelligent buildings. 
\end{abstract}

\section{Introduction}\label{sec:MS} 

Innovative engineering is always looking for adequate instruments to design software systems and support developers all along the development process to deploy correct software. Microservices~\cite{Dragoni2017} recently demonstrated to be an effective architectural paradigm to cope with software complexity, and in particular scalability~\cite{DLLMMS2017}. The success of the paradigm has been demonstrated in a number of domains, including mission-critical systems~\cite{DDLM2017}. 

Around the concept of microservice a number of activities emerged, both of scientific or purely engineering interest. The field of Microservices Science and Engineering (MSE) is not completely established at the moment, and neither it is clearly defined. In this paper, we offer an overview intended as a collection of bibliographic references and links to the field, focusing mostly on recent applications we have been working and on the activities of our team. We aim at focusing on three major aspects: (1) the emerging of the Microservice architectural style and its peculiarities (2) a language-based approach to support Microservice (3) applications, for example in programming languages and intelligent buildings.

The paper is structured as follows. After this short introduction, in Section~\ref{sec:MS} we will discuss the main concepts of Microservice literature. In Section~\ref{sec:Jolie} we will introduce the Jolie programming language, an open source project aimed at supporting microservice development from a linguistic point of view. In Section~\ref{sec:applications} we will discuss the contribution of our research team to the development of the Jolie programming language and in the field of Smart Building. Section~\ref{sec:conclusions} will finally draw some conclusive remarks.

\section{What is a microservice?}\label{sec:MS} 

Microservices~\cite{Dragoni2017} are not just \textit{small services}, which means little by itself. 
It is an architectural style that originated from Service-Oriented Architectures (SOAs)~\cite{mackenzie2006} \cite{sillitti2002}, that we will try to emphasize here. The main idea is to move \textit{in the small} (within an application) some of the concepts that worked \textit{in the large}, i.e. for cross-organization business-to-business workflow which makes use of orchestration engines such as WS-BPEL (in turn inheriting some of the functional principles from concurrency theory~\cite{LucchiM07}).

When following the microservice paradigm, a system is structured by composing small independent building blocks communicating exclusively via message passing. These components are called \textit{microservices}. The characteristic differentiating the new style from monolithic architectures and classic Service-Oriented is the emphasis on \textbf{scalability}, \textbf{independence}, and \textit{semantic cohesiveness} of each unit constituting the system. 

\begin{figure}[!ht]
\centering
\includegraphics[width=0.3\textwidth]{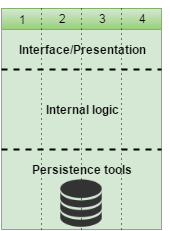}
\caption{Monolith Architecture \label{comparison}}
\end{figure}

Indeed, mainstream languages for development of server-side applications (e.g. Java, C/C++, Python) still provide abstractions to break down the complexity of programs into modules or components \cite{predonzani2001} \cite{clark2004} \cite{gross2005}, but these languages are designed for the creation of single executable artifacts. In monolithic architecture the modularization abstractions rely on the sharing of resources of the same machine (memory, databases, files) and the components are therefore not independently executable. In Figure \ref{comparison}, the classic monolithic organization is pictorially described: here the different layers of the system, from presentation to access to persistence tools, and including the business logic, are split in terms of responsibilities between different modules (here indicated by the vertical split with numbers from 1 to 4). In fact, each module may take part in the implementation of functionalities related to each layer, the database is common, and so the access to other resources such as memory.  

\begin{figure}[!ht]
\centering
\includegraphics[width=0.7\textwidth]{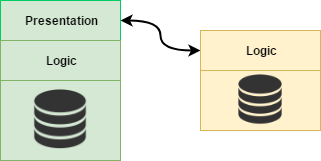}
\caption{Microservices Architecture \label{ms}}
\end{figure}

A notable problem of monoliths is \textit{maintainability} and \textit{evolvability}, all issues related to change. In~\cite{Dragoni2017} a detailed description of these aspects is given, together with our own definition of \textit{microservice} which tries to shed some light in the currently intricate and young literature. Figure \ref{ms} shows how the componentization is done in a microservice architecture: each own service has a dedicated persistence tool and communication is via message passing. In this kind of organization there is no vertical split through all the system layers and the deployment is independent. The complexity is moved to the level of coordination of services (often called orchestration~\cite{Mazzara2005}). Moreover, a number of additional problems need to be addressed due to the distributed nature of this kind of approach (e.g., trust and certification \cite{damiani2009}).

The first set of question asked in this context typically is: \textit{how small?} Is a Microservice a \textit{very small service}? What does it mean? How do we measure size (Line of codes, size of executable, 
number of classes or modules, size of API, size of team)?

A Microservice is not just a \textit{very small service}. There is not a predefined size limit that defines whether a service is a microservice or not. Indeed microservice is a somehow misleading definition. Each microservice is expected to implement a single \textit{business capability}, in fact a very limited system functionality, bringing benefits in terms of service maintainability and extendability. Since each microservice represents a single business capability, which is delivered and updated independently, discovering  bugs or adding minor improvements do not have any impact on other services and on their releases. In common practice, it is also expected that a single service can be developed and managed by a single team~\cite{Dragoni2017}. 

The idea to have a team working on a single microservice is rather appealing: to build a system with a modular and loosely coupled design, one should pay attention to the organization structure and its communication patterns as they, according to Conway's Law~\cite{conway1968committees}, directly impact the produced design. So if one creates an organization with each team working on a single service, such structure will make the communication more efficient not only on the team level, but within the whole organization, improving the resulting design in terms of modularity. Microservices' approach is to keep teams small and communications efficient by creating small cross-functional (DevOps) teams that are able to continuously work on the same service and to be fully responsible for it (the ``you build it, you run it'' principle~\cite{gray2006conversation}). The teams are organized around services, which in turn are organized around business capabilities~\cite{ms_article} The optimal team size for microservices is best described by Jeff Bezos’ famous ``two pizza team'' rule, which suggests that the size of a team should be no larger than what two pizzas can feed. The rule itself does not give an exact number, however it is possible to estimate it to be around 6-8 people. The drawback of such approach is that it is not always practical from the financial point of view to maintain a dedicated team of developers for a single service as it may lead to high development/maintenance costs~\cite{jones2014soa}. Furthermore, one should be careful when designing the high level structure of the organization using microservices - increasing  the number of services will negatively impact  on the overall organization efficiency, if no actions are taken. 

The second set of questions that often arises is instead: \textit{is this the same story than SOA?} What are the differences? Indeed there are some notable differences. In SOA, services are not required to be self-contained with data and User Interface, and their own persistence tools, eg. database. SOA has no focus on independent deployment units and related consequences, it is simply an approach for business-to-business intercommunication. The idea of SOA was to enable business-level programming through business processing engines and languages such as WS-BPEL and BPMN that were built on top of the vast literature on business modelling~\cite{YanMCU07}. Furthermore, the emphasis was all on \textit{service orchestration} more than service development and deployment.

Microservices have seen their popularity blossoming with an explosion of concrete applications seen in real-life software~\cite{N15}. Several companies are involved in a major refactoring of their back-end systems in order to improve scalability \cite{DLLMMS2017}. In~\cite{DDLM2017} a real world case study, concerning the migration of a mission critical system from an existing monolithic architecture to microservices, has been presented. This case study shows the will of major companies to cope with scalability issues.

\section{Jolie: a Language-based Approach}\label{sec:Jolie}

The notable success of the approach gave rise to both academic and commercial interest, and ad-hoc programming languages arose to address the new architectural style \cite{MGZ14}. In principle, any general-purpose language could be used to program microservices. However, some of them are more oriented towards scalable applications and concurrency \cite{Guidi2017} . The Jolie (Java Orchestration Language Interpreter Engine)~\cite{MGZ14} programming language, for example, is based on the new paradigm and it allows describing computation from a data-driven instead of process-driven perspective \cite{Safina2016}. As another advantage, Jolie has already a large community of users and developers \cite{Bandura16}.

Jolie is a functional programming language that combines a multiplicity of aspects that are destined to revolution the way in which software is conceived, designed and understood. Originated from a major formalization effort~\cite{sensoria} for workflow and service composition~\cite{Mazzara11}, the language does not integrate a notion of correctness; it is simply built on it. The intuitiveness of the message-passing paradigm supports the design phase and avoids side effects that are not trivial to test. Four important concepts are identified to be first class entities in the programming language in order to address the microservice architecture:

\begin{enumerate}
\item \textit{Interfaces}: to support modular programming, services has to be deployed as \textit{black boxes}. In order to compose services in larger systems, interfaces have to describe the provided functionalities and those required from the environment.
\item \textit{Ports}: since a microservice interacts with other services, a communication port describes how its functionalities are made available to the network (interface, communication technology, and data protocol). Ports should be specified separately from the implementation of a service. Input ports describe the functionalities that the service provides to the rest of the system, while output ports describe the functionalities that the service requires from the rest of the system.
\item \textit{Workflows}: structured protocols appear repeatedly in microservices and they are not natively supported by mainstream languages. All possible operations are always enabled (for example in Object-Oriented programming). Causal dependencies are programmed by using a book-keeping variable, which is error-prone, and it does not scale when the number of causality links increases. A microservice language should  provide abstractions for programming workflows.
\item \textit{Processes}: workflows define the blueprint of the behavior of a service. At runtime a service may interact with multiple clients and other external services, therefore there is need to support multiple concurrent executions of its workflow. A process is a running instance of a workflow, and a service may include many processes executing concurrently. Each process runs independently of the others, to avoid interference, and has its own private state.
\end{enumerate}

Let us illustrate the Jolie syntax with a simple example of the service printing anything it receives. First we need to define the interface that other services will use and list all available functions inside (as depicted in Figure \ref{jol1}).
\begin{figure}[H]
\centering
 {\normalsize
  \[
  \begin{array}{l}
    \jolkw{interface} \jol{ PrintInterface \{}\\
	\hspace*{1cm}\jolkw{OneWay}\jol{: print (}\jolkw{ string }\jol{)}\\
    \jol{\}}
  \end{array}
  \]
}
\caption{interface code \label{jol1}}
\end{figure}
\vspace{-1.5cm}
This interface declares the \texttt{one-way} function \jol{PrintInterface}, meaning that any service using this interface will be able to call or provide this function without receiving or, correspondingly, providing the response. Then we define the printing service itself, listing the service entry point's name (\jol{PrintService}), location, protocol and interfaces it uses (see Figure \ref{jol3}). The behavior is described in the \jol{main} part of the service. The behavior is composed of the one function \jol{print}, printing the line it receives.
\begin{figure}[H]
\centering
\vspace{-1.5cm} 
 {\normalsize
  \[
  \begin{array}{l}
  	\jolkw{include} \jol{ ``console.iol'' }\\\\
  	\jolkw{include} \jol{ ``printInterface.iol'' }\\\\
    \jolkw{outputPort} \jol{ PrintService \{}\\
	\hspace*{1cm}\jolkw{Location}\jol{: ``socket://localhost:8000''}\\
	\hspace*{1cm}\jolkw{Protocol}\jol{: json}\\
	\hspace*{1cm}\jolkw{Interfaces}\jol{: printInterface}\\
    \jol{\}}\\ \\
    \jol{main \{}\\
    \hspace*{1cm}\jol{print( line )\{}\\
    \hspace*{1.5cm}\jol{print@Console( line )()}\\
    \jol{\}}
  \end{array}
  \]
}
\caption{Server's code \label{jol3}}
\end{figure}
\vspace{-1cm}
Finally, we define the client's service, including the information needed for calling the printing service and call to the printing function (\jol{print@PrintService})
\begin{figure}[H]
\centering

  {\normalsize
  \[
  \begin{array}{l}
  	\jolkw{include} \jol{ ``printInterface.iol'' }\\\\
    \jolkw{outputPort} \jol{ PrintService \{}\\
	\hspace*{1cm}\jolkw{Location}\jol{: ``socket://localhost:8000''}\\
	\hspace*{1cm}\jolkw{Protocol}\jol{: json}\\
	\hspace*{1cm}\jolkw{Interfaces}\jol{: printInterface}\\
    \jol{\}}\\ \\
    \jol{main \{}\\
    \hspace*{1cm}\jol{print@PrintService(``Hello, world!'')}\\
    \jol{\}}
  \end{array}
  \]
}
 
\caption{Client's code \label{jol2}}
\end{figure}
\vspace{-0.7cm}
After invoking both services, \jol{PrintService} will print our ``Hello, world!'' greetings.

Jolie is an open source project with an active community of developers. Our team has been working on an extension of the type system~\cite{Safina2016} and the development of static type checking with refinement types~\cite{Tchitchigin16}, as well as development of the IDE~\cite{Bandura16}.
One of the current projects relates to the augmenting of user experience. We are trying to make the language easy to use, adding the inline documentation, value scaffolding, autocompletion and other ergonomics improving features.  

However, there are more ongoing projects aimed on ensuring Jolie type safety. The approach is to implement the type checker from \cite{mingela2017} follows the formal specification rules defined in~\cite{nielsen}. The rules then are encoded on the Jolie interpreter level and checked by means of Z3 SMT solver~\cite{MouraB08}. \cite{akentev2017} follows a slightly different approach, it is 
built on top of a proof assistant instead of a SMT solver, which helps to ascertain the correctness of the specification. The type checker is written as well-typed program  by means of dependent types in Agda~\cite{agda} programming language.

From the architectural point of view, Jolie has the potential to lead to a paradigm shift. Component-wise each building block is built as a microservice~\cite{M16} embedding business capabilities in isolation. Every microservice can be reused, orchestrated, and aggregated with others~\cite{montesi}. This approach brings simplicity in components management, reducing development and maintenance costs, and supporting distributed deployments~\cite{fowler-tradeoffs}.

\section{Applications in Smart Buildings}\label{sec:applications}  
 
The ideal application scenario where scalability, minimality and cohesiveness demonstrate their effectiveness is Smart Buildings. There are several different devices on the market that have been used in the Internet of Things and smart buildings-related projects. None of these projects, however, was so far developed using the Jolie programming language.

\begin{figure}[!ht]
\centering
\includegraphics[width=0.9\textwidth]{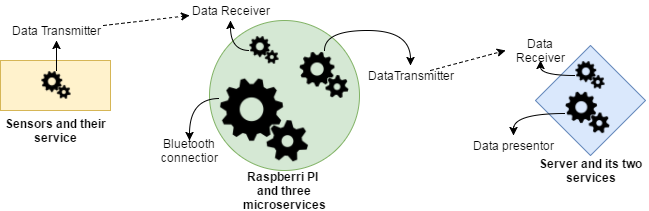}
\caption{Sensors infrastructure architecture \label{infrastructure}}
\end{figure}

Our team at Innopolis has developed an infrastructure of sensors in the University building \cite{Salikhov2016b,Salikhov2016a}. This solution allows to monitor an equipped area and therefore collect data that can be mined and analyzed for specific purposes. The system is taking advantage of the Jolie programming language to coordinate nodes and user interface. The nodes used in this system consist of Raspberry Pi micro-computers~\cite {Raspberri}, Texas Instruments Sensor Tags~\cite{TIST}, door sensor and a web camera.  Currently, this system is able to collect and analyze room temperature, pressure and illumination level. It is also able to distinguish and count people, which are located in the covered area. Figure~\ref{infrastructure} shows the general project infrastructure where each sensor has a related service to transmit data, the Raspberry Pi micro-computer is running services responsible to receive and transmit data to the server, and the server presents the data.

The future plan is to design and realize an automatic Personal Assistant which is capable to observe the data, learn about different users preferences, and adapt the room conditions accordingly for the different phases of his/her work. To develop this, it will be necessary to operate speech and visual recognition via machine learning, and connect these functionalities to the existing system.

\vspace{-0.8cm}
\section{Conclusions}\label{sec:conclusions} 

There is no free lunch someone said. Indeed, a Microservice architecture is, in general, more complex than one based on monolith. This is the cost of growing and scaling easily. Despite of this, companies of considerable size are migrating their mission critical systems  (of considerable size) into the new architectural style demonstrating an early understanding of how critical scalability is, and how costs would differently grow later~\cite{DDLM2017}.

In this paper, we presented the basic principles of Service Science and Engineering (SSE), with the applications developed by our research team. We also supported the idea that a language-based approach seems the best choice to cope with microservice development. Summarizing, the following are the significant advantages of microservices: (1) Smaller code base therefore simpler to develop, test, deploy, scale (2) easier for new developers and it allows fast start (3) Polyglot architecture (each service may use individual technology) (4) Evolutionary design (remove, add, replace services).

We are actively collaborating with both the scientific world (to develop solid theories and methodologies in order to improve software quality) and with companies interested to migrate their systems. The next decade will see a growing attention to the SSE field, and the development of further programming languages intended to address the paradigm. Changes to scene should be expected, and these may be comparable to what Object-Oriented programming brought in the last two decades of the previous century.

\vspace{-0.2cm}
\bibliographystyle{ieeetr}
\bibliography{references}

\end{document}